\documentstyle[11pt]{article}
\newcounter{subequation}[equation]

\makeatletter 
 
\def\thesubequation{\theequation\@alph\c@subequation}
\def\@subeqnnum{{\rm (\thesubequation)}}
\def\slabel#1{\@bsphack\if@filesw {\let\thepage\relax
   \xdef\@gtempa{\write\@auxout{\string
      \newlabel{#1}{{\thesubequation}{\thepage}}}}}\@gtempa
   \if@nobreak \ifvmode\nobreak\fi\fi\fi\@esphack}
\def\subeqnarray{\stepcounter{equation}
\let\@currentlabel=\theequation\global\c@subequation\@ne
\global\@eqnswtrue
\global\@eqcnt\z@\tabskip\@centering\let\\=\@subeqncr
$$\halign to \displaywidth\bgroup\@eqnsel\hskip\@centering
  $\displaystyle\tabskip\z@{##}$&\global\@eqcnt\@ne
  \hskip 2\arraycolsep \hfil${##}$\hfil
  &\global\@eqcnt\tw@ \hskip 2\arraycolsep
  $\displaystyle\tabskip\z@{##}$\hfil
   \tabskip\@centering&\llap{##}\tabskip\z@\cr}
\def\endsubeqnarray{\@@subeqncr\egroup
                     $$\global\@ignoretrue}
\def\@subeqncr{{\ifnum0=`}\fi\@ifstar{\global\@eqpen\@M
    \@ysubeqncr}{\global\@eqpen\interdisplaylinepenalty \@ysubeqncr}}
\def\@ysubeqncr{\@ifnextchar [{\@xsubeqncr}{\@xsubeqncr[\z@]}}
\def\@xsubeqncr[#1]{\ifnum0=`{\fi}\@@subeqncr
   \noalign{\penalty\@eqpen\vskip\jot\vskip #1\relax}}
\def\@@subeqncr{\let\@tempa\relax
    \ifcase\@eqcnt \def\@tempa{& & &}\or \def\@tempa{& &}
      \else \def\@tempa{&}\fi
     \@tempa \if@eqnsw\@subeqnnum\refstepcounter{subequation}\fi
     \global\@eqnswtrue\global\@eqcnt\z@\cr}
\let\@ssubeqncr=\@subeqncr
\@namedef{subeqnarray*}{\def\@subeqncr{\nonumber\@ssubeqncr}\subeqnarray}
\@namedef{endsubeqnarray*}{\global\advance\c@equation\m@ne%
                           \nonumber\endsubeqnarray}

\makeatletter
\@addtoreset{equation}{section}
\makeatother
\renewcommand{\theequation}{\thesection.\arabic{equation}}

\def\dalemb#1#2{{\vbox{\hrule height .#2pt
        \hbox{\vrule width.#2pt height#1pt \kern#1pt
                \vrule width.#2pt}
        \hrule height.#2pt}}}

\let\a=\alpha \let\b=\beta   \let\e=\epsilon
  \let\q=\theta  
  \let\n=\nu   
     
       \let\D=\Delta

\def\nn{\nonumber} \def\bd{\begin{document}} \def\ed{\end{document}}
\def\ds{\documentstyle} \let\fr=\frac \let\bl=\bigl \let\br=\bigr
\let\Br=\Bigr \let\Bl=\Bigl 
\let\bm=\bibitem
\let\na=\nabla
\let\pa=\partial \let\ov=\overline
\def\ie{{\it i.e.\ }} 
\newcommand{\be}{\begin{equation}} 
\newcommand{\ee}{\end{equation}} 
\def\ba{\begin{array}}
\def\ea{\end{array}}
\def\ft#1#2{{\textstyle{{\scriptstyle #1}\over {\scriptstyle #2}}}}
\def\fft#1#2{{#1 \over #2}}
\def\del{\partial}
\def\sst#1{{\scriptscriptstyle #1}}
\def\oneone{\rlap 1\mkern4mu{\rm l}}
\def\e7{E_{7(+7)}}
\def\td{\tilde}
\def\wtd{\widetilde}
\def\im{{\rm i}}
\def\bog{Bogomol'nyi\ }
\def\q{{\tilde q}}
\def\hast{{\hat\ast}}
\def\0{{\sst{(0)}}}
\def\1{{\sst{(1)}}}
\def\2{{\sst{(2)}}}
\def\3{{\sst{(3)}}}
\def\4{{\sst{(4)}}}
\def\5{{\sst{(5)}}}
\def\6{{\sst{(6)}}}
\def\7{{\sst{(7)}}}
\def\8{{\sst{(8)}}}
\def\n{{\sst{(n)}}}
\def\oo{{\"o}}
\def\hA{\hat{\cal A}}
\def\ns{{\sst {\rm NS}}}
\def\rr{{\sst {\rm RR}}}
\def\tH{{\widetilde H}}
\def\tB{{\widetilde B}}
\def\cA{{\cal A}}
\def\cF{{\cal F}}
\def\tF{{\wtd F}}
\def\Z{\rlap{\sf Z}\mkern3mu{\sf Z}}
\def\ep{{\epsilon}}
\def\IIA{{\rm IIA}}
\def\IIB{{\rm IIB}}
\def\ads{{\rm AdS}}
\def\R{\rlap{\rm I}\mkern3mu{\rm R}}

\def\Ei{{\hbox{Ei}}}
\def\Ci{{\hbox{Ci}}}
\def\Si{{\hbox{Si}}}

\newcommand{\ho}[1]{$\, ^{#1}$}
\newcommand{\hoch}[1]{$\, ^{#1}$}
\newcommand{\bea}{\begin{eqnarray}} 
\newcommand{\eea}{\end{eqnarray}} 
\newcommand{\ra}{\rightarrow}
\newcommand{\lra}{\longrightarrow}
\newcommand{\Lra}{\Leftrightarrow}
\newcommand{\ap}{\alpha^\prime}
\newcommand{\bp}{\tilde \beta^\prime}
\newcommand{\tr}{{\rm tr} }
\newcommand{\Tr}{{\rm Tr} } 
\newcommand{\NP}{Nucl. Phys. }
\newcommand{\tamphys}{\it Center for Theoretical Physics,
Texas A\&M University, College Station, TX 77843}
\newcommand{\upenn}{\it Dept. of Phys. and Astro., 
University of Pennsylvania,
Philadelphia, PA 19104}

\newcommand{\auth}{M. Cveti\v{c}\hoch{1}, H. L\"u\hoch{1}, 
J. F. V\'azquez-Poritz}

\begin{document}
\begin{flushright}
UPR/0846-T \\
April 1999\\
\hfill{\bf hep-th/9904135}\\
\end{flushright}


\begin{center}
{\large {\bf  Massive-scalar Absorption by Extremal $p$-branes}
}

\vspace{20pt}

\auth

\vspace{10pt}
{\hoch{\dagger}\upenn}

\vspace{30pt}

\underline{ABSTRACT}
\end{center}

          We study the absorption probability of minimally-coupled
massive scalars by extremal $p$-branes.  In particular, we find that
the massive scalar wave equation under the self-dual string background
has the same form as the massless scalar wave equation under the
dyonic string background. Thus it can be cast into the form of a
modified Mathieu equation and solved exactly.  Another example that we
can solve exactly is that of the $D=4$ two-charge black hole with
equal charges, for which we obtain the closed-form absorption
probability.  We also obtain the leading-order absorption
probabilities for D3-, M2- and M5-branes.

{\vfill\leftline{}\vfill
\vskip 10pt \footnoterule {\footnotesize \hoch{1} Research supported
in part by DOE grant DOE-FG02-95ER40893
\vskip  -12pt} \vskip   14pt
}

\pagebreak
\setcounter{page}{1}


\section{Introduction\label{sec:intro}}

       There has been considerable interest recently in studying
absorption probabilities for fields propagating in various black hole
and $p$-brane backgrounds \cite{dmw1,gk1,mast,cgkt,km,kleb,gkt,cl1,%
cl2,emp,ghkk,lm,mm,gubhash,taylor,ah,clpt,clpt2}.  One of the motivations
is the conjectured duality of supergravity on an AdS spacetime and the
conformal field theory on the boundary of the AdS
\cite{mald,gkp,witten}.  Previously, the study of absorption was
mainly concentrated on the case of massless scalars.  Some work has
been done for the cases of the emission of BPS particles from five-
and four-dimensional black holes \cite{gk1}.  These BPS particles can
be viewed as pp-waves in a spacetime of one higher dimension. Hence
they satisfy the higher-dimensional massless wave equations.

         In this paper, we consider the absorption probability of
minimally-coupled massive particles by extremal $p$-branes. The wave
equation for such a scalar depends only on the metric of the
$p$-brane, which has the form
\be
ds^2= \prod_{\a=1}^N H_\a^{-\ft{\td d}{D-2}}\,
dx^\mu dx^\nu \eta_{\mu\nu}
+ \prod_{\a=1}^N H_\a^{\ft{d}{D-2}}\, dy^mdy^m
\ ,\label{dmetric}
\ee
where $d=p+1$ is the dimension of the world volume of the $p$-brane,
$\td d =D-d-2$, and $H_\a= 1 + Q_\a/r^{\td d}$ are harmonic functions
in the transverse space $y^m$, where $r^2 =y^m\, y^m$. (Note that the
ADM mass density and the physical charges of the extreme $p$-brane
solutions are proportional to $\sum_{i=1}^NQ_i$ and $Q_i$,
respectively)

 It follows that the wave equation, $\del_{\sst M} (\sqrt{g}\,
g^{\sst{MN}}\del_{\sst N} \Phi)=m^2\ \Phi$, for the massive
minimally-coupled scalar, with the ansatz, $\Phi(t, r, \theta_i)
=\phi(r)\, Y(\theta_i)\, e^{-\im \omega t}$, takes the following form:
\be
\fft{d^2\phi}{d\rho^2} + \fft{\td d+1}{\rho}\, \fft{d\phi}{d\rho} +
\Big[
\prod_{\a=1}^N (1 +\fft{\lambda^{\td d}_\a}{\rho^{\td d}})
-\fft{\ell(\ell +\td d)}{\rho^2} -
\fft{m^2}{\omega^2} 
\prod_{\a=1}^N (1+\fft{\lambda_\a^{\td d}}{\rho^{\td d}}
)^{\ft{d}{D-2}} \Big]\, \phi =0\ ,\label{genwave}
\ee
where $\rho=\omega\, r$ and $\lambda_\a = \omega\, Q_\a^{1/\td d}$.
Note that when $m=0$, the wave equation depends on $\td d$, but is
independent of the world-volume dimension $d$.  This implies that the
wave equation for minimally-coupled massless scalars is not invariant
under the vertical-dimensional reduction, but is invariant under
double-dimensional reduction of the corresponding $p$-brane
\cite{clpt2}.  However, for massive scalars, the wave equation
(\ref{genwave}) is not invariant under either double or vertical
reductions.

        The absorption probability of massless scalars is better
understood.  It was shown that for low frequency the
cross-section/frequency relation for a generic extremal $p$-brane
coincides with the entropy/temperature relation of the near extremal
$p$-brane \cite{clpt2}.  There are a few examples where the wave
equations can be solved exactly in terms of special functions.
Notably, the wave equations for the D3-brane \cite{gubhash} and the
dyonic string \cite{clpt} can be cast into modified Mathieu equations.
Hence, the absorption probability can be obtained exactly, order by
order, in terms of a certain small parameter.  There are also examples
where the absorption probabilities can be obtained in closed-form for
all wave frequencies \cite{clpt2}.

         When the mass $m$ is non-zero, we find that there are two
examples for which the wave function can be expressed in terms of
special functions and, thus, the absorption probabilities can be
obtained exactly.  One example is the wave equation in the self-dual
string background, which can be cast into a modified Mathieu
equation. Therefore, we can obtain the exact absorption probability,
order by order, in terms of a certain small parameter.  We discuss
this example in section 2.  Another example is the wave equation for
the $D=4$ two-charge black hole with equal charges.  The wave function
can be expressed in terms of Kummer's regular and irregular confluent
hypergeometric functions.  It follows that we can obtain the
absorption probability in closed-form, which we present in section 3.
In both of the above examples, the massive scalar wave equation has
the same form as the massless scalar wave equation under the
backgrounds where the two charges are generically non-equal.

         However, in general, the massive scalar wave equation
(\ref{genwave}) cannot be solved analytically.  For low-frequency
absorption, the leading-order wave function can be obtained by
matching wave functions in inner and outer regions.  In section 4, we
make use of this technique to obtain the leading-order absorption
probability for D3-, M2- and M5-branes.

\section{Massive scalar absorption for the self-dual string}

     For the self-dual string ($Q_1=Q_2\equiv Q$), we have $d=\td d=2$
and $\lambda_1 =\lambda_2 \equiv\lambda =\omega \sqrt{Q}$.  It follows
that the wave equation (\ref{genwave}) becomes
\be
\fft{d^2\phi}{d\td \rho^2} +
\fft{3}{\td \rho}\, \fft{d\phi}{d\td \rho} +
\Big(1 + \fft{\td\lambda_1^2 + \td\lambda_2^2 -
\ell(\ell+2)}{\td\rho^2}
+\fft{\td\lambda_1^2\,\td\lambda_2^2}{\td \rho^4}\Big)
\, \phi =0\ ,\label{mselfdualstring}
\ee
where
\bea
&&\td \lambda_1 = \lambda\ ,\qquad \td \lambda_2 = \lambda\, \sqrt{
1-(m/\omega)^2}\ ,\nn\\
&&\td \rho= \rho\, \sqrt{1-(m/\omega)^2}\ ,
\eea  
Thus the wave equation of a minimally-coupled massive scalar on a
self-dual string has precisely the same form as that of a
minimally-coupled massless scalar on a dyonic string, where $\td
\lambda_1$ and $\td \lambda_2$ are associated with electric and
magnetic charges.  It was shown in \cite{clpt} that the wave equation
(\ref{mselfdualstring}) can be cast into the form of a modified
Mathieu equation, and hence the equation can be solved exactly.  To do
so, one makes the following definitions
\be
\phi(\td \rho) = \fft{1}{\rho}\, \Psi(\rho)\, \qquad
\td \rho = \sqrt{\td\lambda_1\, \td\lambda_2}\, e^{-z}\ .
\ee
The wave equation (\ref{mselfdualstring}) then becomes the modified
Mathieu equation \cite{clpt}
\be
\Psi'' + (8\Lambda^2\, \cosh(2z) - 4\a^2)\, \Psi =0\ ,
\ee
where
\bea
\alpha^2 &=& \ft14(\ell+1)^2 -\Lambda^2\, \Delta\ ,\nn\\
\Lambda^2 &=& \ft14 \td\lambda_1\, \td\lambda_2 =
\ft14 \lambda^2\, \sqrt{1 -(m/\omega)^2}\ =\ft14
\omega\sqrt{\omega^2-m^2}Q,\label{aldpara}\\
\Delta &=& \fft{\td\lambda_1}{\td\lambda_2} +
\fft{\td\lambda_2}{\td\lambda_1} =\sqrt{1-(m/\omega)^2}
+\fft1{\sqrt{1-(m/\omega)^2}}\ .
\eea

       The Mathieu equation can be solved, order by order, in terms of
$\Lambda^2$.  The result was obtained in \cite{clpt}, using the
technique developed in \cite{doug}. (For an extremal D3-brane, which
also reduces to the Mathieu equation, an analogous technique was
employed in \cite{gubhash}.)  In our case there are two parameters,
namely $\omega R$ and $m/\omega$.  We present results for two
scenarios:

\subsection{Fixed mass/frequency ratio probing}

        In this case, we have $m/\omega=\beta$ fixed.  The requirement
that $\Lambda$ is small is achieved by considering low-frequency and, 
hence, small mass of the probing particles.  In this case, $\Delta$ is
fixed, and the absorption probability has the form \cite{clpt}
\be
P_\ell =  \fft{4\pi^2\, \Lambda^{4+4\ell}}{(\ell+1)^2 \,
\Gamma(\ell+1)^4}\, \sum_{n\ge0}
\sum_{k=0}^n b_{n,k}\,
\Lambda^{2n}\,  (\log \bar\Lambda)^k\ ,\label{lamexp}
\ee
where $\bar\lambda = e^\gamma\, \lambda$, and $\gamma$ is Euler's
constant. The prefactor is chosen so that $b_{0,0}=1$.  Our results
for the coefficients $b_{n,k}$ with $k\le n\le 3$ for the first four
partial waves, $\ell=0,1,2,3$, were explicitly given in \cite{clpt}.
In particular the result up to the order of $\Lambda^2$ is given by
\cite{clpt}
\be
P_\ell = \fft{4\pi^2\, \Lambda^{4+4\ell}}{(\ell+1)^2\, \Gamma(\ell+1)^4}
\, \Big[ 1 -\fft{8\D}{\ell+1}\, \Lambda^2\, \log\Lambda +
\fft{4\D\, \Lambda^2}{(\ell+1)^2}\, \Big(1+ 2(\ell+1)\,
\psi(\ell+1)\Big) + \cdots\Big] \ ,
\ee
where $\psi(x)\equiv \Gamma'(x)/\Gamma(x)$ is the digamma function.

\subsection{Fixed mass probing}

         Now we consider the case where the mass of the test particle
is fixed.  In this case, it is ensured that $\Lambda$ is small by
considering the limiting frequency of the probing particle, namely
$\omega \rightarrow m^+$, i.e. the particle is non-relativistic.  In
this limit, the value of $\Delta$ becomes large (while at the same
time the expansion parameter $\Lambda$ can still be ensured to remain
small). Furthermore, we shall consider a special slice of the
parameter space where $\a^2$, given in (\ref{aldpara}), is fixed.  The
absorption probability for fixed $\a$ was obtained in \cite{clpt}.  It
is of particular interest to present the absorption probability for
$\a\to 0$, given by 
\be 
P= \fft{\pi^2}{\pi^2+ (2\log\bar\Lambda)^2}\,
\Big(1 -\ft{32}{3}\Lambda^4\, (\log\bar\Lambda)^2 -\ft{16}{3}\,
(4\zeta(3)-3) \, \fft{\Lambda^4\, \log\bar\Lambda}{ \pi^2+
(2\log\bar\Lambda)^2} +{\cal O}(\Lambda^8) \Big)\ ,\label{a0} 
\ee
where $\bar\Lambda= e^\gamma\, \Lambda$.  When $\a^2<0$, we define
$\a^2 = {\rm i}\,\beta$, and find that the absorption probability
becomes oscillatory as a function of $\Lambda$, given by \cite{clpt}
\be
P= \fft{\sinh^2 2\pi\b}{\sinh^2 2\pi\b + \sin^2(\theta-4\b\,
\log\Lambda)} +\cdots \ ,
\ee
where
\be
\theta={\rm arg}\, \fft{\Gamma(2\im\, \b)}{\Gamma(-2\im\, \b)}\ .
\ee
Note that the $\alpha\to 0$ limit is a dividing domain between the
region where the absorption probability has power dependence of
$\Lambda$ ($\a^2> 0$) and the region with oscillating behavior on
$\Lambda$ ($\a<0$).

\section{Closed-form absorption for the $D=4$ two-charge black hole}

For a $D=4$ black hole, specified in general by four charges $Q_1$,
$Q_2$, $P_1$ and $P_2$~\cite{cvyoum}, we have $d=\td d=1$. We consider
the special case of two equal non-zero charges ($Q_1=Q_2 \equiv Q$
with $P_1=P_2=0$) and therefore $\lambda_1= \lambda_2
\equiv\lambda=\omega Q$.  It follows that the wave equation
(\ref{genwave}) becomes
\be
\frac{d^2 \phi}{d\td \rho^2} + 
\frac{2}{\td \rho}\, \frac{d\phi}{d\td \rho} +
\Big[ \big( 1 + \frac{\td \lambda_1}{\td \rho}\big) \big( 1 +
\frac{\td \lambda_2}{\td \rho}\big) - \frac{\ell (\ell +1)}{\td \rho^2}
\Big]\, \phi =0\, \label{m2chargebh}
\ee
where
\bea
&&\td \lambda_1 = \frac{\lambda}{\sqrt{1-(m/\omega)^2}}\ ,\qquad
\td \lambda_2 = \lambda\, \sqrt{1-(m/\omega)^2}\ ,\nn\\
&&\td \rho= \rho\, \sqrt{1-(m/\omega)^2}\ , 
\eea   
Thus the wave equation of a minimally-coupled massive scalar on a
$D=4$ black hole with two equal charges has precisely the same form as
that of a minimally-coupled massless scalar on a $D=4$ black hole with
two different charges. The closed-form absorption probability for the
latter case was calculated in \cite{clpt2} (see also \cite{balalar}).
The absorption probability for the former case is, therefore, given by
\be
P^{(\ell)}=\frac{1-e^{-2\pi \sqrt{4\lambda^2-(2\ell+1)^2}}}{1+e^{-\pi 
(2\lambda  + \sqrt{4\lambda^2-(2\ell+1)^2})} e^{-\pi \delta}}\ ,\ \ \ \ \
\ \lambda \ge \ell + \frac{1}{2}\ ,
\ee
where
\be
\delta \equiv \td \lambda_1 + \td \lambda_2 - 2\lambda
=\lambda\big[(1-({m/ \omega})^2)^{1/4}-(1-(m/\omega)^2)^{-1/4}\big]^2\ge 0\ ,
\ee
with $P^{(\ell)}=0$ if $\lambda \le \ell + \frac{1}{2}$. In the
non-relativistic case ($\omega\to m^+$), the absorption probability
takes the (non-singular) form $P^{(\ell)}=1-e^{-2\pi
\sqrt{4\lambda^2-(2\ell+1)^2}}$, with $\lambda\sim mQ$.

The total absorption cross-section is given by:
\be
\sigma^{(abs)} = \sum_{\ell\le\lambda-\ft12} 
\frac{\pi \ell(\ell+1)}{\omega^2-m^2}
P^{(\ell)} 
\ee
It is oscillatory with respect to the dimensionless parameter, $M
\omega\sim Q\omega=\lambda$. ($M$ is the ADM mass of the black hole.)
This feature was noted in \cite{sanchez} for Schwarzschild black holes
and conjectured to be a general property of black holes due to wave
diffraction. Probing particles feel an effective finite potential
barrier around black holes, inside of which is an effective potential
well. Such particles inhabit a quasi-bound state once inside the
barrier. Resonance in the partial-wave absorption cross-section occurs
if the energy of the particle is equal to the effective energy of the
potential barrier. Each partial wave contributes a 'spike' to the
total absorption cross-section, which sums to yield the oscillatory
pattern. As the mass of the probing particles increases, the amplitude
of the oscillatory pattern of the total absorption cross-section
decreases.

\section{Leading-order absorption for D3, M2 and M5-branes}

       In the previous two sections, we considered two examples for
which the massive scalar wave equations can be solved exactly.  In
general, the wave function (\ref{genwave}) cannot be solved
analytically.  In the case of low frequency, one can adopt a
solution-matching technique to obtain approximate solutions for the
inner and outer regions of the wave equations.  In this section, we
shall use such a procedure to obtain the leading-order absorption
cross-sections for the D3, M2 and M5-branes.

       We now give a detailed discussion for the D3-brane, for which
we have $D=10$, $d=\td d=4$ and $N=1$.  We define $\lambda \equiv
\omega R$. It follows that the wave equation (\ref{genwave}) becomes
\be
\Big( \frac{1}{\rho^5} \frac{\partial}{\partial \rho} \rho^5 
\frac{\partial}{\partial \rho} + 1 + 
\frac{(\omega R)^4}{\rho^4} - \sqrt{1 + \frac{(\omega
R)^4}{\rho^4}} (\frac{m}{\omega})^2 - 
\frac{\ell (\ell + 4)}{\rho^2} \Big) \phi (\rho) = 0. \label{D3eqn}
\ee
Thus, we are interested in absorption by the Coulomb potential in 6
spatial dimensions. For $\omega R \ll 1$ we can solve this problem by
matching an approximate solution in the inner region to an approximate
solution in the outer region.  To obtain an approximate solution in
the inner region, we substitute $\phi = \rho^{- 3/2} f$ and find
that
\be
\Big( \frac{\partial^2}{\partial \rho^2} + \frac{2}{\rho} 
\frac{\partial}{\partial \rho} - \big( \frac{15}{4} + \ell (\ell + 4) 
\big) \frac{1}{\rho^2} + 1 +
\frac{(\omega R)^4}{\rho^4} - \sqrt{1 + 
\frac{(\omega R)^4}{\rho^4}} \big( \frac{m}{\omega} 
\big) ^2 \Big) f = 0. \label{inner}
\ee
In order to neglect $1$ in the presence of the $\frac{1}{\rho^2}$
term, we require that
\be
\rho \ll 1\ . \label{condition1}
\ee
In order for the scalar mass term to be negligible in the presence of
the $\frac{1}{\rho^2}$ term, we require that
\be
\rho \ll \Big[ \big( \frac{\omega}{m} \big)^4 \big( \frac{15}{4} +
\ell (\ell +4)\big)^2 - (\omega R)^4\Big]^{1/4} \ . \label{condition2}
\ee
Physically we must have $m \le \omega$. Imposing the low-energy
condition $\omega R \ll 1$ causes (\ref{condition1}) to be a stronger
constraint on $\rho$ than is (\ref{condition2}).  Under the above
conditions, (\ref{inner}) becomes
\be
\Big( \frac{\partial^2}{\partial \rho^2} + 
\frac{2}{\rho} \frac{\partial}{\partial \rho} - 
\big( \frac{15}{4} + \ell (\ell + 4) \big) \frac{1}{\rho^2} + 
\frac{(\omega R)^4}{\rho^4} \Big) f = 0\ , \label{approxinner}
\ee
which can be solved in terms of cylinder functions. Since we
are interested in the incoming wave for $\rho \ll 1$, the appropriate
solution is
\be
\phi_o = i \frac{(\omega R)^4}{\rho^2} \Big( J_{\ell + 2} 
(\frac{(\omega R)^2}{\rho}) + i N_{\ell + 2} 
(\frac{(\omega R)^2}{\rho}) \Big), \ \ \ \ \rho
\ll 1,
\ee
where $J$ and $N$ are Bessel and Neumann functions.  In order to
obtain an approximate solution for the outer region, we substitute
$\phi = \rho^{-5/2} \psi$ into (\ref{D3eqn}) and obtain
\be
\Big( \frac{\partial^2}{\partial \rho^2} - \big( \frac{15}{4} + 
\ell (\ell + 4) \big) \frac{1}{\rho^2} + 1 + 
\frac{(\omega R)^4}{\rho^4} - \sqrt{1 + \frac{(\omega
R)^4}{\rho^4}} (\frac{m}{\omega})^2 \Big) \psi = 0. \label{outer}
\ee
In order to neglect $\frac{(\omega R)^4}{\rho^4}$ in the presence of
the $\frac{1}{\rho^2}$ term, we require that
\be
\rho \gg (\omega R)^2\ . \label{condition3}
\ee
Within the scalar mass term, $\frac{(\omega R)^4}{\rho^4}$ can be
neglected in the presence of 1 provided that
\be
\rho \gg \omega R\ . \label{condition4} 
\ee
Imposing the low-energy condition, $\omega R \ll 1$, causes
(\ref{condition4}) to be a stronger constraint on $\rho$ than
(\ref{condition3}).  Under the above conditions, (\ref{outer}) becomes
\be
\Big( \frac{\partial^2}{\partial \rho^2} - \big( \frac{15}{4} + 
\ell (\ell + 4) \big) \frac{1}{\rho^2} + 1 - (\frac{m}{\omega})^2 \Big) 
\psi = 0. \label{approxouter}
\ee
Equation (\ref{approxouter}) is solved in terms of cylinder functions:
\be
\phi_{\infty} = A \rho^{- 2} J_{\ell + 2} (\sqrt{1 - (m/\omega)^2} \rho)
+ B \rho^{- 2} N_{\ell + 2} (\sqrt{1 - (m/\omega)^2} \rho), 
\ \ \ \ \rho \gg \omega R,
\ee
where $A$ and $B$ are constants to be determined.

Our previously imposed low-energy condition, $\omega R \ll 1$, is
sufficient for there to be an overlapping regime of validity for
conditions (\ref{condition1}) and (\ref{condition4}), allowing the
inner and outer solutions to be matched. Within the matching region,
all cylinder functions involved have small arguments. We use the same
asymptotic forms of the cylinder functions as used by \cite{clpt2}.
We find that $B=0$ and
\be
A = \frac{4^{\ell+2} \Gamma (\ell+3) \Gamma (\ell+2)}{\pi 
\big( 1-(\frac{m}{\omega})^2 \big)^{\frac{\ell+2}{2}} 
(\omega R)^{2\ell}}
\ee
The absorption probability is most easily calculated in this
approximation scheme as the ratio of the flux at the horizon to the
incoming flux at infinity.  In general, this flux may be defined as
\be
F = i \rho^{\td d +1} \big( \bar{\phi} 
\frac{\partial \phi}{\partial \rho} - \phi 
\frac{\partial \bar{\phi}}{\partial \rho} \big)\ ,
\ee
where $\phi$ here is taken to be the in-going component of the
wave. From the approximate solutions for $\phi$ in the inner and outer
regions, where the arguments of the cylinder functions are large, we
find that the in-going fluxes at the horizon and at infinity are given
by
\be
F_{horizon}=\frac{4}{\pi} \omega^4 R^8,\ \ \ \ \ \ \ \ \ \ 
F_{\infty}=\frac{A^2}{\pi \omega^4}\ .
\ee
Thus, to leading order, the absorption probability, $P \equiv
F_{horizon}/F_{\infty}$, is
\be
P^{(\ell)} = \frac{\pi^2 \big( 1-\big( \frac{m}{\omega} \big)^2 
\big)^{\ell + 2} (\omega R)^{4\ell+8}}{4^{2\ell+3} 
(\ell+2)^2 [(\ell+1)!]^4}
\ee
In general, the phase-space factor relating the absorption probability
to the absorption cross-section can be obtained from the massless
scalar case considered in \cite{unruh} with the replacement $\omega
\rightarrow \sqrt{\omega^2-m^2}$:
\be
\sigma^{(\ell)}=2^{n-2}\pi^{n/2-1}\Gamma(n/2-1)(\ell+n/2-1) 
{\ell+n-3 \choose \ell} (\omega^2-m^2)^{(1-n)/2} P^{(\ell)}
\ee
where $n=D-d$ denotes the number of spatial dimensions.
Thus, for the D3-brane we find
\be
\sigma_{3-brane}^{(l)} = \frac{\pi^4 (\ell+3)(\ell+1)
[1-(m/\omega)^2]^{\ell-1/2}}{(3)2^{4\ell+3}[(\ell+1)!]^{4}} 
\omega^{4l+3} R^{4l+8}\ .
\ee
As can be seen, within our approximation scheme, the effects of a
nonzero scalar mass amount to an overall factor in the partial
absorption cross-section. Also, the s-wave absorption cross-section is
increased by $m$ and the higher partial wave absorption cross-sections
are diminished by $m$. This is to be expected, since the scalar mass
serves to increase gravitational attraction as well as rotational
inertia.

The above approximation scheme can be applied to massive scalar
particles in all $N=1$  $p$-brane backgrounds except for the case
of
$D=11$  $p$-branes with $\td d=4$ and $\td d=5$, in which cases the
scalar mass term cannot be neglected in the inner region. For $N >
1$, we are unable to find solvable approximate equations which give
an overlapping inner and outer region.

For the M2-brane, we have $D=11$, $d=3$, $\td d =6$ and $N=1$:
\be
\sigma_{\rm M2-brane}^{(l)} = \frac{\pi^5(\ell+5)(\ell+4)
[1-(m/\omega)^2]^{\ell-1/2}}{(15)2^{3\ell+2}\ell!(\ell+2)! 
\Gamma^2(\frac{3+\ell}{2})} \omega^{3\ell+2} R^{3\ell+9}
\ee
For the M5-brane, we have $D=11$, $d=6$, $\td d =3$ and $N=1$: 
\be
\sigma_{\rm M5-brane}^{(l)} = \frac{2^{2\ell+5} \pi^3 
(\ell+2)(\ell+3/2)(\ell+1)[(\ell+1)!]^2 
[1-(m/\omega)^2]^{\ell-1/2}}{(2\ell+3)^2[(2\ell+2)!]^4}
\omega^{6\ell+5} R^{6\ell+9}
\ee

In fact, for all $N=1$ $p$-branes, other than the two for which the
approximation scheme cannot be applied, the partial absorption
cross-sections have the same additional factor due to the scalar mass:
\be
\sigma_{\rm massive}^{\ell}=
\sigma_{\rm massless}^{\ell} [1-(m/\omega)^2]^{\ell-1/2},
\ee
for $m \le \omega$, and $\sigma_{\rm massless}^\ell$ has the same form as
the leading-order absorption for massless scalars.  Note that the 
suppression [enhancement] of the partial
cross-section for $\ell\ge 1$ [for $\ell =0$], when the
non-relativistic limit is taken.

\section{Conclusions}

In this paper we have addressed the absorption cross-section for
minimally-coupled massive particles in the extreme $p$-brane
backgrounds. In particular, we found exact absorption probabilities in
the cases of the extreme self-dual dyonic string in $D=6$ and two
equal-charge extreme black hole in $D=4$. Notably these two examples
yield the same wave equations as that of the minimally coupled
massless scalar in the $D=6$ extreme dyonic string, and two charge
$D=4$ extreme black hole backgrounds, respectively. Namely, one of the
two charge parameters in the latter (massless) case is traded for the
scalar mass parameter in the former (massive) case.  Thus, for these
equal charge backgrounds, the scattering of minimally-coupled massive
particles can be addressed explicitly, and the distinct behavior of
the absorption cross-section on the energy $\omega$ (or equivalently
momentum $p\equiv \sqrt{\omega^2-m^2}$) is studied.  In particular,
the non-relativistic limit of the particle motion gives rise to a
distinct, resonant-like absorption behavior in the case of the
self-dual dyonic string.

We have also found corrections due to the scalar mass for the
leading-order absorption cross-sections for D3-, M2- and M5-branes. In
particular, in the non-relativistic limit, there is the expected
suppression [enhancement] in the absorption cross-section for partial
waves $\ell\ge 1$ [$\ell=0$].

The results obtained for the absorption cross-section of the
minimally-coupled massive scalars, in particular those in the extreme
self-dual dyonic string background, may prove useful in the study of
AdS/CFT correspondence~\cite{mald}. Namely, the near-horizon region of
the extreme dyonic string background has the topology of $AdS_3\times
S^3$, with the $AdS_3$ cosmological constant $\Lambda$ and the radius
$R$ of the three-sphere ($S^3$) related to the charge $Q$ of the
self-dual dyonic string as $\Lambda= R^2= \sqrt{Q}$ (see e.g.,
\cite{cvlar3}).  On the other hand, the scattering of the
minimally-coupled massive fields (with mass ${\cal M}$) in the $AdS_3$
background yields information~\cite{gkp,witten} on the correlation
functions of the operators of the boundary $SL(2,{\bf R})\times
SL(2,{\bf R})$ conformal field theory~\cite{hen} with conformal
dimensions $h_{\pm}=\ft12(1\pm\sqrt{1+{\cal
M}^2\Lambda^2})$~\cite{bala2}. The scattering analyzed here
corresponds to that of a minimally-coupled massive scalar in the the
full self-dual string background, rather than in only the truncated
$AdS_3$ background.  These explicit supergravity results may, in turn,
shed light on the pathologies of the conformal field theory of the
dyonic string background~\cite{seiwitt}.

\end{document}